\documentclass[usenatbib,letters]{mnras}
\usepackage{amsmath} % 
\usepackage{amssymb} % 
\usepackage{bm} % 
\usepackage[pdftex]{graphicx} % 

\def\apjl{ApJ}
\def\apjs{ApJS}
\def\ao{Appl.~Opt.}

\def\aap{A\&A}

\newcommand{\ofe}{\text{[O/Fe]}}
\newcommand{\feh}{\text{[Fe/H]}}
\newcommand{\teff}{T_{\mathrm{eff}}}
\newcommand{\lgg}{\log{g}}
\newcommand{\lgeps}{\log\epsilon_{\mathrm{O}}}
\newcommand{\corr}{\Delta\lgeps}
\newcommand{\rom}[1]{\uppercase\expandafter{\romannumeral #1\relax}}
\newcommand{\triplet}{\text{O\rom{1} 777\,nm}}
\newcommand{\forbidden}{\text{[O\rom{1}] 630\,nm}}
\newcommand{\uvresonance}{\text{O\rom{1} 130\,nm}}
\newcommand{\fig}[1]{\text{Fig.\,\ref{#1}}}
\newcommand{\sect}[1]{\text{\S\ref{#1}}}
\newcommand{\stagger}{\textsc{stagger}}
\newcommand{\mtd}{\textsc{multi3d}}
\newcommand{\scate}{\textsc{scate}}

\title[The Galactic chemical evolution of oxygen]{The Galactic chemical evolution of oxygen inferred from 3D non-LTE spectral line formation calculations}
\author[A.~M.~Amarsi, M.~Asplund, R.~Collet, and J.~Leenaarts]{A.~M.~Amarsi$^{1}$\thanks{E-mail: anish.amarsi@anu.edu.au}
, M.~Asplund$^{1}$, R.~Collet$^{1}$, and J.~Leenaarts$^{2}$\\
$^{1}$ Research School of Astronomy and Astrophysics,
Australian National University, ACT 2611, Australia\\
$^{2}$ Institute for Solar Physics, 
Stockholm University, AlbaNova University Centre, 
SE-10691 Stockholm, Sweden}
\date{Accepted 2015 August 19.  Received 2015 July 22; in original form 2015 June 18.}
\pagerange{\pageref{firstpage}--\pageref{lastpage}} \pubyear{---}
\begin{document}
\maketitle 
\label{firstpage}
%-------------------------------------------------------------------------------
\begin{abstract}
We revisit the Galactic chemical evolution of oxygen,
addressing the systematic errors inherent in 
classical determinations of the oxygen abundance
that arise from the use of  
one dimensional hydrostatic (1D) model atmospheres
and from the assumption of local thermodynamic equilibrium (LTE).
We perform detailed 3D non-LTE radiative transfer 
calculations for atomic oxygen lines 
across a grid of 3D hydrodynamic \stagger~model
atmospheres for dwarfs and subgiants.
We apply our grid of predicted line strengths 
of the \forbidden~and \triplet~lines using
accurate stellar parameters from the literature.
We infer a steep decay in \ofe~for $\feh\gtrsim-1.0$,
a plateau $\ofe\approx0.5$ down to 
$\feh\approx-2.5$ and an increasing trend
for $\feh\lesssim-2.5$. 
Our 3D non-LTE calculations yield
overall concordant results from the two 
oxygen abundance diagnostics. 

\end{abstract}
%-------------------------------------------------------------------------------
\begin{keywords}
radiative transfer --- line: formation --- stars: 
atmospheres --- stars: abundances --- Galaxy: abundances --- methods: numerical
\end{keywords}
%-------------------------------------------------------------------------------

\section{Introduction}
\label{introduction}

Oxygen abundances and abundance ratios
are crucial for understanding 
the evolution of stars and galaxies. 
It is therefore disconcerting that 
the \ofe\footnote{The logarithmic abundance
of an arbitrary element A is defined with respect to hydrogen:
$\log\epsilon_{\text{A}}=\log\frac{N_{\text{A}}}{N_{\text{H}}}+12$.
The abundance ratio of elements A and B is given by:
$[\text{A}/\text{B}]=(\log\epsilon_{\text{A}}-\log\epsilon_{\text{A}}^{\odot})-(\log\epsilon_{\text{B}}-\log\epsilon_{\text{B}}^{\odot})$,
where $\odot$ denotes the solar value.}~vs \feh~relationship
in the metal-poor halo is still in dispute,
in spite of tremendous effort over a number of years
\citep[for an in-depth review, see for example][]{2012EAS....54.....S_short}.

The situation is complicated by 
the availability of different oxygen abundance diagnostics,
that often yield discordant results.
Abundances inferred from the 
low-excitation forbidden \forbidden~line 
tend to show a plateau at $\ofe\approx0.5$ below $\feh\approx-1$ 
\citep{2002A&amp;A...390..235N,2004A&amp;A...416.1117C,
2006A&amp;A...451..621G},
as do those inferred 
from OH infrared (IR) vibration rotation lines
\citep{2001NewAR..45..529B,2002ApJ...575..474M}.
Analyses of the high-excitation permitted \triplet~lines 
\citep{2000A&amp;A...356..238C,2002A&amp;A...390..235N,2006A&amp;A...451..621G}
and of the OH ultraviolet (UV) excitation lines
\citep{1998ApJ...507..805I,2001ApJ...551..833I,1999AJ....117..492B,2010A&amp;A...519A..46G}
tend to predict increasing oxygen abundances
toward lower metallicities,
with $\ofe\approx0.7$ at $\feh\approx-2.0$
and $\ofe\approx1.0$ at $\feh\approx-3.0$.
Notably, however, \citet{2006A&amp;A...451..621G} 
inferred a plateau at $\ofe\approx0.5$ from the OH UV lines
in metal-poor subgiants between $\feh=-1$ and $-3$.
It has been suggested \citep[e.g.][]{2001ApJ...551..833I}
that these different trends could reflect 
in part different rates of circumstellar mixing in dwarfs,
subgiants and giants, 
because at low metallicities the different diagnostics are 
typically used in different types of stars. 
\citet{2005A&amp;A...430..655S} compared abundances 
in giants which have
and have not undergone internal mixing.
They found no significant systematic
difference in \ofe,
indicating that the problem must in fact arise from elsewhere.

The discrepancies between the
\forbidden~line and the \triplet~lines 
likely have at least two points of origin.
Firstly, classical abundance analyses
are prone to systematic modelling errors.
These errors arise from the use of one dimensional (1D)
hydrostatic model atmospheres
and from the assumption of local thermodynamic equilibrium (LTE).
Although the \forbidden~line forms in LTE \citep{1995A&amp;A...302..578K},
the effects of inhomogeneities and
three-dimensional structures in late-type
stellar atmospheres on the \forbidden~line
in the metal-poor regime are significant 
\citep{2002A&amp;A...390..235N,2007A&amp;A...469..687C}.
The \triplet~lines are a classic case study 
of non-LTE line formation
\citep{1974A&amp;A....31...23S,1993A&amp;A...275..269K,
2003A&amp;A...402..343T,2009A&amp;A...500.1221F}.
However, full 3D non-LTE calculations have not been 
been carried out beyond the Sun
\citep{1995A&amp;A...302..578K,2004A&amp;A...417..751A,
2009A&amp;A...508.1403P,2013MSAIS..24..111P,
2015arXiv150803487S}.

Secondly, systematic errors can be present in the adopted 
stellar parameters.
In particular, offsets in the adopted effective temperature ($\teff$)
would reveal themselves as offsets in the 
$\ofe$ vs $\feh$ trend, 
as the \forbidden~line and the \triplet~lines
have opposite sensitivities to $\teff$. 
If $\teff$ is underestimated for a given star,
the \forbidden~line will \emph{underestimate} the 
atmospheric oxygen abundance,
while the \triplet~lines will \emph{overestimate} it.
The accurate calibrations of the infrared flux method
(IRFM) by \citet{2010A&amp;A...512A..54C}
have shown previous $\teff$ estimates 
dwarfs and subgiants with $\feh\lesssim-2.5$
to typically be $\sim200\,\mathrm{K}$ too low.

Offsets in $\teff$ of this magnitude are
not sufficient to explain by themselves the \ofe~vs \feh~problem
at the lowest metallicities
\citep{2003ApJ...595.1154F}.
Likewise, extremely large 3D non-LTE abundance corrections 
would be required in the \triplet~lines
to explain the problem, 
because the non-LTE abundance corrections in these lines
\citep{2009A&amp;A...500.1221F,2014A&amp;A...565A.121D}
go in the same direction as
the 3D abundance corrections in the \forbidden~line
\citep{2002A&amp;A...390..235N,2007A&amp;A...469..687C}.
Thus, errors in the stellar parameters must be  
addressed simultaneously with 1D LTE errors to
achieve reliable results from the oxygen abundance diagnostics.

In this letter we aim to pin down 
the true nature of the 
Galactic chemical evolution of oxygen.
To that end, we have 
performed for the first time detailed 3D non-LTE radiative transfer 
calculations for atomic oxygen lines 
across a grid of 3D hydrodynamic 
\stagger~model atmospheres 
\citep{2011JPhCS.328a2003C,2013A&amp;A...557A..26M}
(\sect{method}).
We illustrate the results in the form of 
3D non-LTE vs 1D LTE abundance corrections 
(\sect{results}).
Using our 3D non-LTE grid,
accurate stellar parameters from the literature,
and a large collection of published equivalent widths
of the \forbidden~and \triplet~lines,
we revisit the \ofe~vs \feh~relationship
(\sect{gce}).

%-------------------------------------------------------------------------------
\section{Calculations}
\label{method}

\begin{figure}
\begin{center}
\includegraphics[scale=0.32]{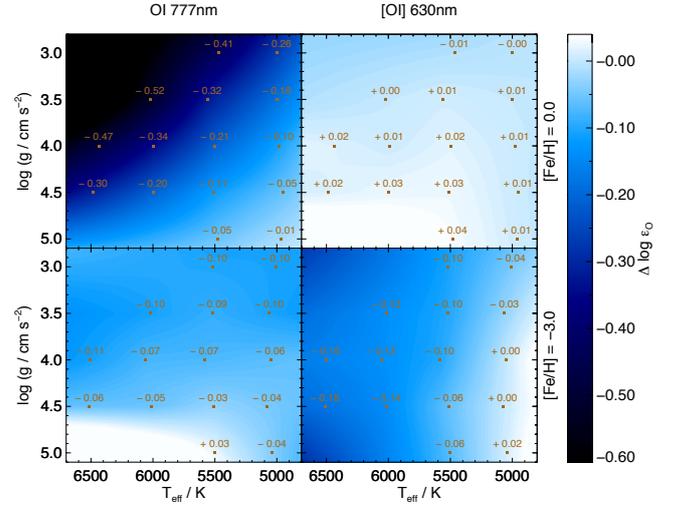}
\caption{Abundance corrections across the \stagger~grid
of 3D stellar atmospheres.
Values beyond the periphery of the grid 
were obtained by linear extrapolation and should be 
treated with caution.
Abundance corrections are shown across 
$\teff$-$\lgg$ planes.
\emph{Top: } $\feh=0.0$; $\lgeps=8.7$.
\emph{Bottom: }$\feh-3.0$; $\lgeps=6.2$.
The 1D microturbulence is $\xi=1.0\,\mathrm{km\,s^{-1}}$.}
\label{fig1}
\end{center}
\end{figure}

\subsection{Model atmospheres}

3D hydrodynamic model atmospheres
were taken from the \stagger-grid 
\citep{2011JPhCS.328a2003C,2013A&amp;A...557A..26M}.
These were trimmed by removing the very optically thick layers
of the original simulations and resampled:
the meshsize was thus reduced from $240^{2}\times230$ gridpoints
down to $120^{2}\times110$, 
the last dimension representing the vertical.
One-dimensional (1D) hydrostatic model atmospheres
were computed using an equation of state
and a treatment of opacities 
and radiative transfer fully 
consistent with the \stagger~models.

We illustrate the extent of the grid in effective temperature
($\teff$) and surface gravity ($\lgg$) space in \fig{fig1}.
Grid nodes are spaced in intervals 
of $\Delta\lgg=0.5$
and $\Delta\teff\approx500\,\mathrm{K}$.
The grid spans $-3.0\leq\feh\leq0.0$,
in equal intervals of $\Delta\feh=1.0$.

\subsection{Spectral line formation}

Multi-level statistical equilibrium
radiative transfer calculations were
performed using the domain-decomposed MPI-parallelised code
\mtd~\citep{2009ASPC..415...87L}.
For a given oxygen abundance and 3D model,
the average line profile was calculated from
four temporal snapshots.
The monochromatic average non-LTE to LTE ratio was then applied to 
the average line profile from at least 50 temporal snapshots
calculated in LTE using \scate~\citep{2011A&amp;A...529A.158H}.

The 3D and the 1D spectral line formation
calculations were carried out in the same way
to avoid incurring systematic errors.
In 1D an additional parameter, the microturbulence $\xi$, is required to
describe the line broadening by convective velocity fields.
The 1D grid was calculated for several values of $\xi$;
in this letter we adopt a typical value 
of $\xi=\text{1.0}\,\mathrm{km\,s^{-1}}$ 
to illustrate the abundance corrections (\sect{results}).
We emphasise that in 3D no microturbulent velocities
are necessary
\citep{2000A&amp;A...359..729A}.

\subsection{Model atom}

A 23-level model oxygen atom was used,
based on the models used by  
\citet{1993ApJ...402..344C},
\citet{1993A&amp;A...275..269K},
and \citet{2009A&amp;A...500.1221F}.
The model was updated with 
energies and oscillator strengths
from NIST\footnote{http://www.nist.gov/pml/data/asd.cfm} 
\citep{kramida2012nist},
natural line broadening coefficients 
from VALD3\footnote{http://vald.astro.univie.ac.at/$\sim$vald3/php/vald.php} 
\citep{1995A&amp;AS..112..525P,2000BaltA...9..590K},
and collisional line broadening coefficients from
\citet{1998PASA...15..336B}.

The statistical equilibrium
is sensitive to the adopted collisional rate coefficients.
Results from R-matrix calculations were used 
for the rate of excitation via collisions with electrons
\citep{2007A&amp;A...462..781B}.
Following \citet{1993PhST...47..186L},
a correction to the original formula
of \citet{1968ZPhy..211..404D,1969ZPhy..225..483D} 
was adopted for the
rate of excitation via collisions with neutral hydrogen.
All levels were coupled in this manner, including 
including those with radiatively weak and forbidden transitions,
in which case Drawin's formula was 
used with an effective oscillator strength
$f_{\mathrm{min}}=10^{-3}$.
An enhancement factor $S_{\mathrm{H}}$ to the Drawin formula was calibrated
by studying the centre-to-limb variation in the Sun,
following \citet{2009A&amp;A...508.1403P}
using observational data from the Swedish Solar Telescope.
We have obtained $S_{\mathrm{H}}\approx1$
and have inferred a solar oxygen abundance $\lgeps\approx8.68$.

The dominant systematic errors in the non-LTE
calculations likely arise from the errors
in the rate of excitation via neutral hydrogen collisions.
As discussed in \citet{2011A&amp;A...530A..94B},
the approach of calibrating the Drawin formula
using a single parameter $S_{\mathrm{H}}$
cannot correct any errors in the relative rates,
and can mask other deficiencies in the models. 
One can get a feeling for the sensitivity 
of the results on the neutral hydrogen collisions
by comparing the results
with those obtained when $S_{\mathrm{H}}$ is reduced to nought, 
thereby neglecting them altogether.
When doing so, the derived non-LTE abundance corrections for the 
\triplet~can become up to 0.05 dex more negative.

\subsection{Background opacities}
\label{background opacities}
While background continuum opacities were calculated 
internally by the two spectral line formation codes,
background line opacities had to be included separately. 
These were pre-calculated assuming LTE
and the chemical composition of the 
atmosphere being modelled 
\citep{2005A&amp;A...442..643C,2012MNRAS.427...27B}.

It can be tempting to simplify the non-LTE analysis
by neglecting the background lines
that do not overlap with the lines being studied in detail.
This is dangerous; oxygen proves to be a case in point.
We have found the Lyman-$\alpha$ line to play a 
critical role in controlling the statistical equilibrium
of oxygen in the photosphere.
Without the Lyman-$\alpha$ line,
extremely large abundance corrections would 
erroneously be inferred from 
the \triplet~lines in metal-poor stars.
We discuss this briefly in \sect{results},
and in more detail in a forthcoming paper.

The \forbidden~line is heavily blended with a NiI line
\citep{1978MNRAS.182..249L,2001ApJ...556L..63A}.
This NiI line was included in LTE in both 3D and 1D,
with a nickel abundance that was scaled relative 
to the solar value $\log\epsilon_{\mathrm{Ni}}=6.20$
\citep{2015A&amp;A...573A..26S}.

%-------------------------------------------------------------------------------
\section{Abundance corrections}
\label{results}

\begin{figure}
\begin{center}
\includegraphics[scale=0.32]{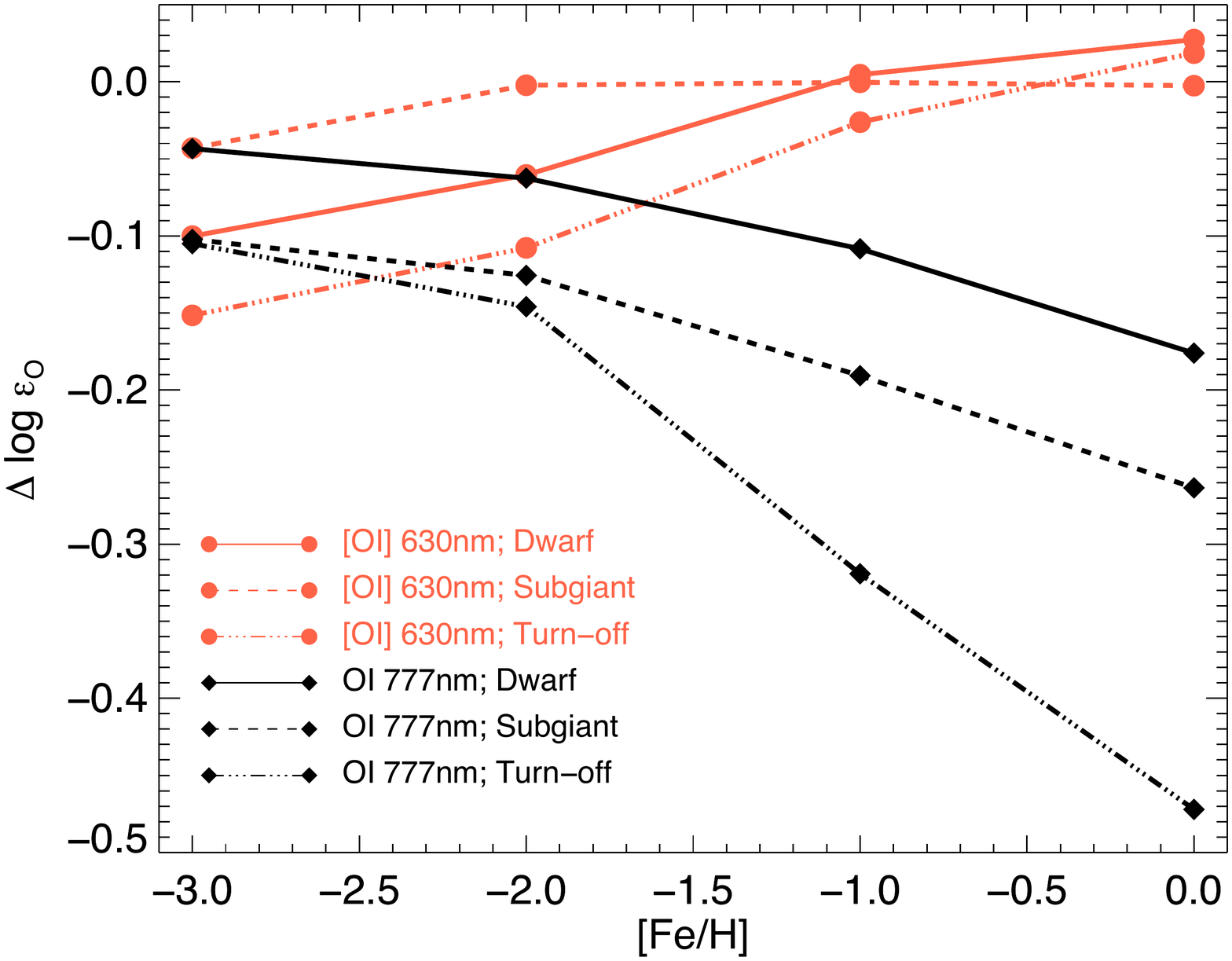}
\caption{Abundance corrections 
$\corr=\lgeps^{\mathrm{3D,NLTE}}-\lgeps^{\mathrm{1D,LTE}}$
for the \forbidden~line
(\emph{red circles}) 
and for the \triplet~lines
(\emph{black diamonds}) 
in a dwarf (\emph{solid}; $\teff=5771\,\mathrm{K}$, $\lgg=4.44$),
a subgiant (\emph{dash}; $\teff=5000\,\mathrm{K}$, $\lgg=3.0$),
and a turn-off star
(\emph{dot-dash}; $\teff=6500\,\mathrm{K}$, $\lgg=4.0$),
with 1D microturbulence
$\xi=1.0\,\mathrm{km\,s^{-1}}$ and varying chemical composition.
The 1D LTE oxygen abundance is
$8.7+\feh+[\alpha/\text{Fe}]$, where the enhancement
is $[\alpha/\text{Fe}]=0.5$ for $\feh\leq-1.0$ and $[\alpha/\text{Fe}]=0.0$ 
for $\feh=0.0$.}
\label{fig2}
\end{center}
\end{figure}

Abundance corrections 
$\corr=\lgeps^{\mathrm{3D,NLTE}}-\lgeps^{\mathrm{1D,LTE}}$
were calculated by demanding that
the equivalent width of the line flux
obtained in LTE from the 1D model
with a given oxygen abundance $\lgeps^{\mathrm{1D,LTE}}$
matches that
inferred in non-LTE from the 3D model
with some oxygen abundance $\lgeps^{\mathrm{3D,NLTE}}$.
Equivalent widths were determined by direct integration
across the line profile. 
We illustrate corrections for the two oxygen 
abundance diagnostics 
across different models in \fig{fig1} and \fig{fig2}.

The 3D corrections for the \forbidden~line
are close to zero in the solar metallicity case, 
but become increasingly more negative at lower \feh,
in line with previous findings
\citep{2002A&amp;A...390..235N,2007A&amp;A...469..687C}.
Differences in the predicted mean temperature stratification 
between the hydrodynamical 3D models and the 
hydrostatic 1D models
are more severe in this regime;
1D models typically overestimate the atmospheric
temperature in the upper layers 
where this line forms \citep{1999A&amp;A...346L..17A}.

In contrast, the 3D non-LTE vs 1D LTE
abundance corrections for the \triplet~lines
are most severe in the solar metallicity case,
reaching up to $\Delta\lgeps\approx-0.5$
(\fig{fig1} and \fig{fig2}).
Photon losses in the line itself 
drive the non-LTE effects
\citep{2004A&amp;A...417..751A};
inhomogeneities in the 3D model atmospheres exacerbate
these effects. 

\citet{2009A&amp;A...500.1221F} 
obtained extremely large non-LTE effects at 
low \feh~in their 1D non-LTE study of the \triplet~lines.
These authors found that, toward lower metallicities, 
photon pumping \citep{2005ARA&amp;A..43..481A}
in the \uvresonance~lines
would drive very large overpopulations with respect to LTE
in the upper level of these lines.
Then, collisional coupling would propagate
this overpopulation onto the lower level of the \triplet~lines.
This would increase the \triplet~line opacity,
strengthening the lines and
causing large negative abundance 
corrections, reaching $\Delta\lgeps\approx-0.5$ 
in metal-poor turn-off stars
(assuming $S_{\mathrm{H}}=1$).
However, the analysis of \citet{2009A&amp;A...500.1221F} 
did not include background opacity from the Lyman-$\alpha$ line,
which is significant because the line is very strong and
the \uvresonance~lines sit on its red wing.
We have found that the Lyman-$\alpha$ line
provides an efficient alternative destruction route for UV photons 
at low \feh~that completely stifles the photon pumping effect 
in the \uvresonance~lines.
Thus, in the absence of an alternative non-LTE mechanism,
we are left with small departures from LTE
in the \triplet~lines at low \feh.

%-------------------------------------------------------------------------------
\section{The Galactic chemical evolution of oxygen}
\label{gce}

\begin{figure}
\begin{center}
\includegraphics[scale=0.32]{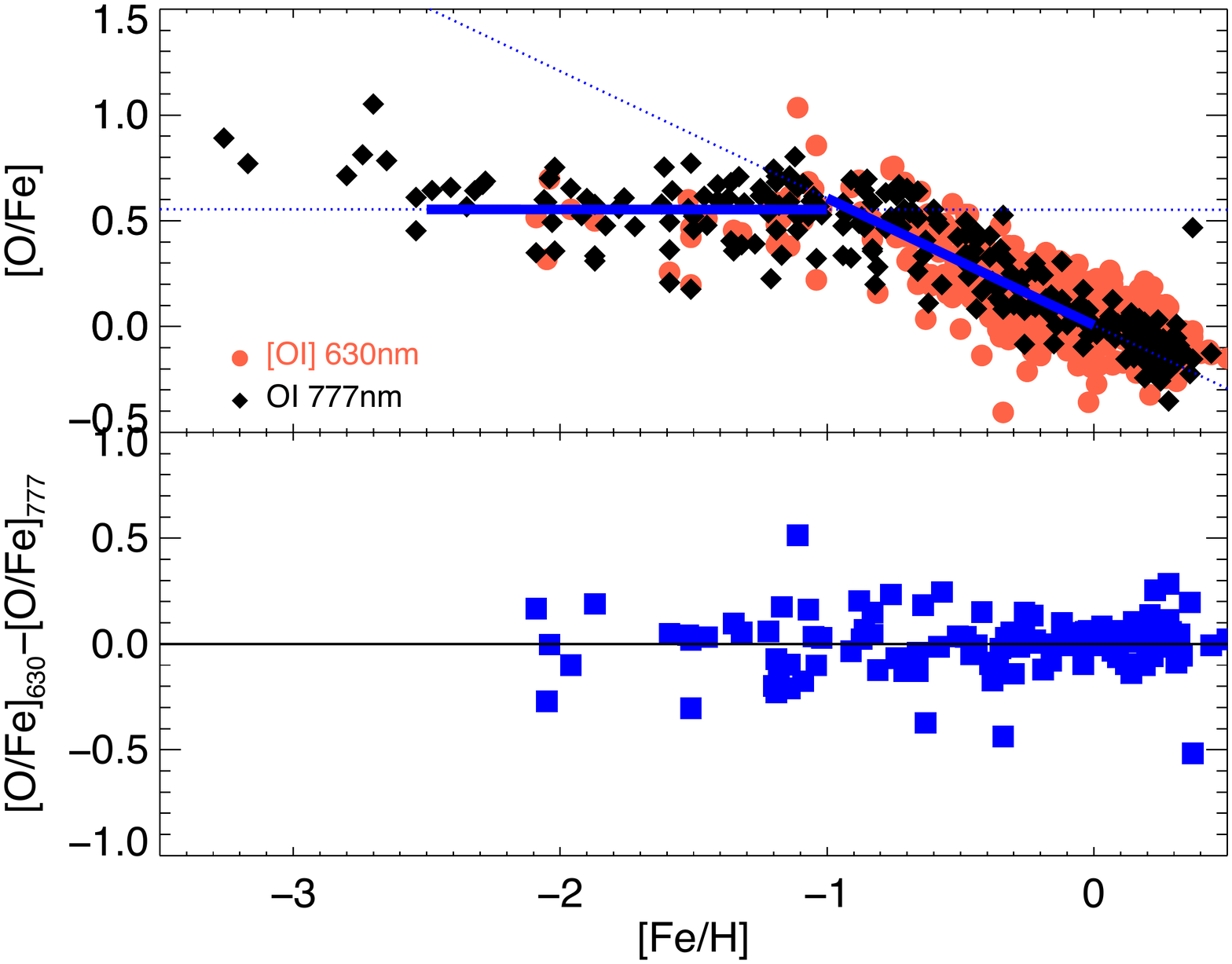}
\caption{Final results based
on equivalent widths from
the literature 
\citep{2000A&amp;A...356..238C,2003ApJ...595.1154F,
2004A&amp;A...415..155B,2004A&amp;A...414..931A,
2006A&amp;A...451..621G,2009A&amp;A...500.1143F,
2002A&amp;A...390..235N,2014A&amp;A...568A..25N,
2015A&amp;A...576A..89B}
and stellar parameters from
\citet{2014A&amp;A...568A..25N}
and \citet{2010A&amp;A...512A..54C,2011A&amp;A...530A.138C}.
\emph{Top:} \ofe~vs \feh~inferred from 
the \forbidden~line (\emph{red circles})
and from the \triplet~lines (\emph{black diamonds}). 
Lines of best fit to the data 
in the domains $-2.5<\feh<-1.0$
and $-1.0<\feh<0.0$ are overdrawn.
The fits were obtained by minimising $\chi^{2}$
and assigning each star equal weight;
where abundances from both diagnostics were available,
the mean value was used.
\emph{Bottom:} differences in \ofe~for individual stars inferred
from the \forbidden~and \triplet~lines,
for the cases where both sets of equivalent widths were available.
The median value is $0.02\,\text{dex}$
and the standard deviation is $0.14\,\text{dex}$.}
\label{fig3}
\end{center}
\end{figure}

The equivalent widths of the \forbidden~and
\triplet~lines were taken from a number of studies
of dwarfs and subgiants
based on high signal-to-noise observations
\citep{2000A&amp;A...356..238C,
2003ApJ...595.1154F,
2004A&amp;A...415..155B,
2004A&amp;A...414..931A,
2006A&amp;A...451..621G,
2009A&amp;A...500.1143F,
2002A&amp;A...390..235N,2014A&amp;A...568A..25N,
2015A&amp;A...576A..89B}.
Where more than one value was available 
for a given star and line,
the unweighted mean was adopted.
For consistency, equivalent widths
across the theoretical 3D non-LTE grid
were computed by fitting Gaussian functions to 
the line fluxes and integrating analytically. 
Stellar parameters were taken from several
recent studies 
\citep{2010A&amp;A...512A..54C,2011A&amp;A...530A.138C,2014A&amp;A...568A..25N}.
These $\teff$ estimates were either derived using
or critically compared to the accurate 
IRFM calibrations of 
\citet{2010A&amp;A...512A..54C};
where more than one set of stellar
parameters was available for a given star,
the newest set was adopted.

Hitherto, studies have typically found
discrepant results from the two abundance diagnostics
at low \feh~(\sect{introduction}). 
In \fig{fig3} we compare the \ofe~ratios inferred
from our analyses.
We have found the \forbidden~line and the \triplet~lines 
to give similar \ofe~vs \feh~trends down to $\feh\approx-2.2$,
the lowest metallicity in which the \forbidden~line is detectable
in halo subgiants and turn-off stars.
Furthermore, the abundances inferred
from these diagnostics in the atmospheres of the same stars
are consistent to within 
a standard deviation of 0.14 dex.

It bears repeating that there are two factors
in our analysis that are absent in most previous studies,
that conspire to give concordant results
between the two abundance diagnostics.
First, we have accounted for 3D non-LTE effects in the \triplet~lines:
these are of decreasing importance towards lower
\feh~, but even then remain significant.
Second, we have used new and accurate stellar parameters:
the more reliable IRFM calibrations give $\teff$ estimates 
that are significantly larger than those typically used in the past.

The \ofe~vs \feh~relationship in \fig{fig3}
reflects the evolution with time of oxygen and iron yields.
Oxygen is synthesised almost entirely
in massive stars ($M\gtrsim8\,M_{\odot}$),
its most abundant isotope $^{16}\text{O}$ being
the endpoint of Helium burning
\citep{2002RvMP...74.1015W,2003hic..book.....C,meyer2008nucleosynthesis}.
Since iron is synthesised in type II supernovae explosions
\citep{2002RvMP...74.1015W},
the plateau at $\ofe\approx0.5$
between $-2.2\lesssim\feh\lesssim-1.0$
indicates that massive stars at this epoch eject an
approximately constant ratio of oxygen and iron upon their deaths,
in agreement with Galactic chemical evolution models 
\citep{2004A&amp;A...421..613F,2006ApJ...653.1145K}.
The steep linear decay seen above $\feh\gtrsim-1.0$
is usually interpreted as 
a sign of the long-lived type Ia supernovae
becoming significant, increasing
the rate of enrichment of iron into the cosmos
\citep{1979ApJ...229.1046T,1997ARA&amp;A..35..503M}.

At $\feh\lesssim-2.5$ there is a slight
upturn in \ofe, but it is
less pronounced than found
from the 1D LTE analyses of OH UV lines by
\citet{1998ApJ...507..805I,2001ApJ...551..833I}
and \citet{1999AJ....117..492B}.
The upturn could indicate a shift in the mass distribution of stars
at earlier epochs;
stars that are more massive
and more metal-poor are expected to yield 
larger oxygen-to-iron ratios \citep{2006ApJ...653.1145K}.
We caution however that inferences in this region
are less reliable because they are based on a very small
sample of stars and on analyses of the \triplet~lines alone.
A larger sample of halo turn-off stars with accurate stellar parameters
and very high signal-to-noise spectra
will be needed to confirm this result.

%-------------------------------------------------------------------------------
\section*{Acknowledgements}
\label{acknowledgements}
AMA and MA are supported by the Australian
Research Council (ARC) grant FL110100012.
RC acknowledges support from the ARC through DECRA grant DE120102940.
This research has made use of the SIMBAD database,
operated at CDS, Strasbourg, France.
This research was undertaken with the 
assistance of resources from the 
National Computational Infrastructure (NCI),
which is supported by the Australian Government.

%-------------------------------------------------------------------------------

\bibliographystyle{mn2e}
\bibliography{/Users/ama51/Documents/work/papers/allpapers/bibl.bib}
%-------------------------------------------------------------------------------

\label{lastpage}
\end{document}